\documentclass[conference]{IEEEtran}
\IEEEoverridecommandlockouts

\usepackage{cite}
\usepackage{amsmath,amssymb,amsfonts}
\usepackage{algorithmic}
\usepackage{graphicx}
\usepackage{textcomp}
\usepackage{xcolor}
\usepackage{hyperref}
\usepackage{url}
\usepackage{booktabs}
\usepackage{listings} 
\usepackage{tabularx}
\usepackage{tcolorbox}

\lstdefinestyle{promptstyle}{
  basicstyle=\small\ttfamily,
  breaklines=true,
  frame=none,
  xleftmargin=0pt,
}

\begin{document}

\title{Evaluating LLMs on Java Code Snippet Adaptation Using a Mutation-Injection Framework}

\author{
\IEEEauthorblockN{Ali Aman}
\IEEEauthorblockA{\textit{University of Windsor} \\
Windsor, Canada \\
burkia@uwindsor.ca}
\and
\IEEEauthorblockN{Muhammad Asaduzzaman}
\IEEEauthorblockA{\textit{University of Windsor} \\
Windsor, Canada \\
masaduzz@uwindsor.ca}
\and
\IEEEauthorblockN{Shaowei Wang}
\IEEEauthorblockA{\textit{University of Manitoba} \\
Winnipeg, Canada \\
shaowei.wang@umanitoba.ca}
\and
\IEEEauthorblockN{Chanchal K. Roy}
\IEEEauthorblockA{\textit{University of Saskatchewan} \\
Saskatoon, Canada \\
croy@cs.usask.ca}
}

\maketitle

\begin{abstract}

\textbf{Background:} Developers frequently reuse code by copying fragments and adapting them to fit new contexts. Existing benchmarks for evaluating large language models (LLMs) on code adaptation either rely on explicit step-by-step instructions, cover only narrow change types such as variable wiring, or operate exclusively at function-level granularity. It remains unknown how well LLMs can adapt code fragments without explicit edit guidance when the required changes are varied and controlled.

\textbf{Objective:} We investigate instruction-free code snippet adaptation in which an LLM must adapt a code fragment to fit its target context without any explicit edit guidance. We study three dimensions: which adaptation types are hardest (RQ1), how performance scales with adaptation complexity (RQ2), and how much surrounding context the model needs (RQ3).

\textbf{Method:} We will construct a dataset of Java code fragments from open-source repositories with strong test coverage and apply a taxonomy of adaptation operators, derived from empirical findings on how developers adapt copied code, using a mutation-injection framework. Working at the code fragment level and controlling the injected changes lets us know exactly what adaptations the model must perform. The unmutated fragment serves as a plausible reference for the changes the model needs to make. LLMs will be evaluated on instruction-free adaptation tasks across three context granularity levels. Correctness will be measured primarily via test-suite re-insertion, complemented by mutation-level inspection.

\end{abstract}


\section{Introduction}
When developers write code, they often stop partway through a function or block and reuse a similar code fragment that already exists, whether elsewhere in the same project, in another repository, or on a site such as Stack Overflow. The developer then needs to adapt the copied code to the current context. This includes renaming variables, swapping APIs, adding error handling, changing control flow, and other similar changes.

In real code reuse, developers do not write out a list of edits. They expect the tool to examine the copied code and the surrounding context, then figure out the necessary changes by itself. The central question is therefore: how effectively can large language models perform \emph{instruction-free} adaptation of code fragments across different types of changes, levels of complexity, and amounts of surrounding context? Throughout this paper we use the term \emph{instruction-free} to mean that the model receives only a generic adaptation directive and no explicit list of required edits, target identifiers, operators, or step-by-step guidance; in other words, no edit-level instructions are provided.

Existing benchmarks and tools come close to this problem but each leaves a meaningful gap. Recent work reports a roughly 20\% performance gap between the easiest and hardest adaptation categories for LLMs even when explicit edit instructions are provided \cite{zhang2026adapteval}, indicating that adaptation type is a real and measurable source of difficulty. However, current benchmarks for code adaptation typically (i) derive tasks from real commits in which multiple change types are entangled, (ii) operate only at function-level granularity, and (iii) are limited in scale by their reliance on hand annotation. Adjacent tools and benchmarks target narrower slices of the problem, such as identifier wiring \cite{liu2023adaptivepaste, wang2025wirl}, prompt-engineering for adaptation \cite{zhang2025instruct}, or repository-level code completion that uses cross-file context but does not target adaptation of copied fragments \cite{liu2023repobench, ding2023crosscodeeval, nguyen2024repoexec}.

Our study addresses this combined gap through three distinguishing contributions. First, we use \emph{mutation injection} to construct adaptation tasks. Starting from real, tested code fragments, we apply adaptation operators in reverse so that the fragment given to the LLM differs from its target form in a known, controlled way. Because we choose which operators to inject, we know exactly which edits the model must perform and can directly measure where it struggles --- something benchmarks built from real commits cannot provide since their change types are entangled. Second, we evaluate at the \emph{code fragment level} rather than only at function level, more closely reflecting how copy-paste actually operates in IDEs. Third, we develop a \emph{taxonomy of adaptation changes} grounded in prior empirical studies and operationalize each entry as a programmable mutation operator, yielding a reusable framework that enables future researchers to evaluate new adaptation tools at scale without manual annotation.

We structure the investigation around three exploratory research questions: which adaptation types are hardest for LLMs without explicit instructions (RQ1), how performance scales with the number and combination of injected changes (RQ2), and how much surrounding context a model needs and whether the optimal amount depends on the adaptation type (RQ3).

\section{Background and Related Work}
\textbf{Code Snippet Adaptation.} Zhang et al.\ analyzed real-world adaptations of online code examples and identified recurring patterns such as variable renaming, API substitution, and logic customization \cite{zhang2019analyzing, zhang2024howdevadapt}, confirming that adaptation is inherently context-dependent and rarely consists of a single change type; these empirical findings form the basis of the taxonomy we use in this study. Early tools automated narrow slices of this process: Jigsaw resolved simple identifier mismatches \cite{cottrell2008jigsaw} and APIzation converted Stack Overflow snippets into reusable method signatures \cite{terragni2021apization}, but both were limited to narrow classes of changes and required significant manual effort or domain-specific rules. Despite these efforts, fully automated support for adaptation remains limited, leaving the broader problem of instruction-free, multi-type adaptation at fragment level without a controlled, scalable evaluation framework.

\textbf{LLMs for Code Editing and Adaptation.} LLMs have shown promise on a wide range of code-related tasks, including generation, repair, and editing \cite{jimenez2024swebench, guo2024codeeditorbench, chi2025editbench}, and several recent benchmarks specifically target adaptation. AdaptEval \cite{zhang2026adapteval} evaluates 164 real-world instances mined from commits, but operates at function-level granularity with entangled change types and limited scale due to manual construction, making systematic fine-grained analysis difficult. AdaptivePaste \cite{liu2023adaptivepaste} uses dataflow-aware pre-training and a parallel-decoder architecture to achieve strong results on variable renaming and identifier wiring, but does not address broader structural adaptations such as API swaps, error handling, or logic modifications. WIRL \cite{wang2025wirl} focuses narrowly on context-aware variable wiring using LLM-based agents, while Instruct or Interact \cite{zhang2025instruct} emphasises prompt engineering strategies rather than systematic evaluation of different change types. Repository-level benchmarks such as RepoBench, CrossCodeEval, and RepoExec \cite{liu2023repobench, ding2023crosscodeeval, nguyen2024repoexec} demonstrate the importance of surrounding context but do not evaluate adaptation of copied code fragments.

\textbf{Mutation-Based Evaluation and Clone Detection.} Roy and Cordy introduced a mutation-injection framework that automatically generates large numbers of known clones by applying editing operators to real code fragments; their taxonomy of clone-creation edits has been widely used to evaluate clone detectors with known ground truth and without manual validation \cite{roy2009mutation, bellon2007comparison, sajnani2016sourcerercc}. We adapt this paradigm to a different goal: applying operators in reverse to produce fragments that an LLM must adapt back toward the original, giving us direct control over which edits the model must perform.

\section{Datasets}
\label{sec:datasets}
We will select a set of mature open-source Java repositories from GitHub. Specifically, we plan to query GitHub for Java repositories with at least 500 stars, a buildable Maven (\texttt{pom.xml}) configuration, a line coverage of at least 70\% as reported by JaCoCo \cite{jacoco}, and at least 50 methods. The star and coverage thresholds serve as proxies for repository maturity and test-suite quality respectively; the Maven requirement is a practical constraint so that our framework can compile each project and execute its test suite through a uniform harness. Exact threshold values will be validated during initial dataset construction and adjusted if they yield too few qualifying repositories. Consistent with filtering pipelines used in comparable Java/Maven benchmarks~\cite{may2026freshbrew}, we conservatively estimate 50--150 qualifying repositories; these thresholds may be adjusted during the pilot phase if they yield too few qualifying repositories, as noted in Section~\ref{sec:execution}, with exact counts reported in the final paper. To avoid domain bias, the selected repositories will span at least five distinct application domains (e.g., web frameworks, data processing, CLI tools, testing frameworks, and utility libraries), assessed using the GitHub topic tags of each repository. Extending the harness to Gradle, Ant, Bazel, and other build systems is left as a follow-up. We note that restricting to Java and Maven is a deliberate scoping decision; the framework itself is designed to be extensible to other languages and build systems, as discussed in Section~\ref{sec:threats}.

From these repositories we will extract candidate code fragments to serve as ground-truth seeds, using Spoon \cite{pawlak2016spoon} to parse each compilation unit and identify fragment boundaries. A candidate fragment is defined as any syntactically complete, contiguous block of 3--20 statements that (a) is covered by at least one test case and (b) contains at least two distinct non-keyword identifiers (computed against the Java reserved-word list). Fragments shorter than three statements are discarded as too small to carry meaningful adaptation signal. A fragment can be a complete method or a smaller sub-method unit such as a loop body, conditional branch, try-catch block, synchronized block, or lambda body --- any contiguous block with a clear syntactic boundary.

\section{Mutation Injection Framework}
\label{sec:methodology}
\label{sec:framework}
We will design a controlled, scalable framework resting on three principles: (1) start from real, working code in open-source repositories with strong test coverage; (2) construct adaptation tasks by applying a taxonomy of adaptation operators \emph{in reverse} via mutation injection, so that the required changes are known and controlled; and (3) evaluate correctness by re-inserting each model output into the original location and re-running the existing test suite. Functional correctness is complemented by mutation-level inspection, which checks which specific operators the model successfully reversed.

\subsection{Adaptation Taxonomy}

We will construct a taxonomy of adaptation operators grounded in prior empirical studies of how developers adapt copied code, including Java-focused analyses of Stack Overflow / GitHub adaptation pairs \cite{zhang2019analyzing}, an LLM-era catalogue of real developer adaptations \cite{zhang2026adapteval}, and interview-based studies of context-driven adaptation \cite{zhang2024howdevadapt}. We further draw on prior work on Stack Overflow snippet completeness, which identifies the structural omissions (undeclared identifiers, missing receivers, missing guards) that motivate many real adaptations \cite{treude2017understanding}. The mutation-injection framework of Roy and Cordy \cite{roy2009mutation} provides the methodological basis: each taxonomy entry is operationalised as a programmable, reversible mutation.

Our taxonomy is \emph{operationally defined}, not exhaustive. Prior catalogues list dozens of adaptation types observed in real developer behaviour \cite{zhang2019analyzing, zhang2026adapteval}; ours is a curated subset of what we can \emph{control and measure} under the constraints of our framework. We may expand this set during the pilot phase if we observe additional frequent change types that satisfy our four inclusion criteria. Each operator must: (i) correspond to a frequently observed adaptation in the empirical literature; (ii) be implementable as a deterministic reverse-mutation pass in Spoon \cite{pawlak2016spoon}; (iii) produce a behavioural difference detectable by the existing test suite when re-inserted (semantics-preserving mutations are filtered out automatically, as described in Section~\ref{sec:pipeline}); and (iv) contribute to a difficulty range spanning local identifier edits to broader structural changes, so that RQ1 yields a meaningful ordering rather than a ceiling or floor. We will incorporate an open-coding phase into our execution pipeline to identify and evaluate additional frequent adaptation patterns not yet captured by the current operators; all additions will be reported in the final paper.

Because our unit of evaluation is the code fragment rather than the enclosing method, the taxonomy is scoped to adaptations that live inside the fragment boundary. Method-signature changes (parameter insertion, deletion, or reordering; return-type changes; \texttt{throws}-clause changes; method renames), file-level changes (imports, top-level annotations), and transformations that convert free-floating statements into a callable method \cite{terragni2021apization} are therefore out of scope. Real copy-paste in IDEs frequently operates on sub-function fragments \cite{kim2004ethnographic}, and adaptation behaviour at this granularity may differ from whole-function adaptation in ways that current benchmarks do not surface.

\begin{figure*}[t]
  \centering
  \includegraphics[width=\textwidth]{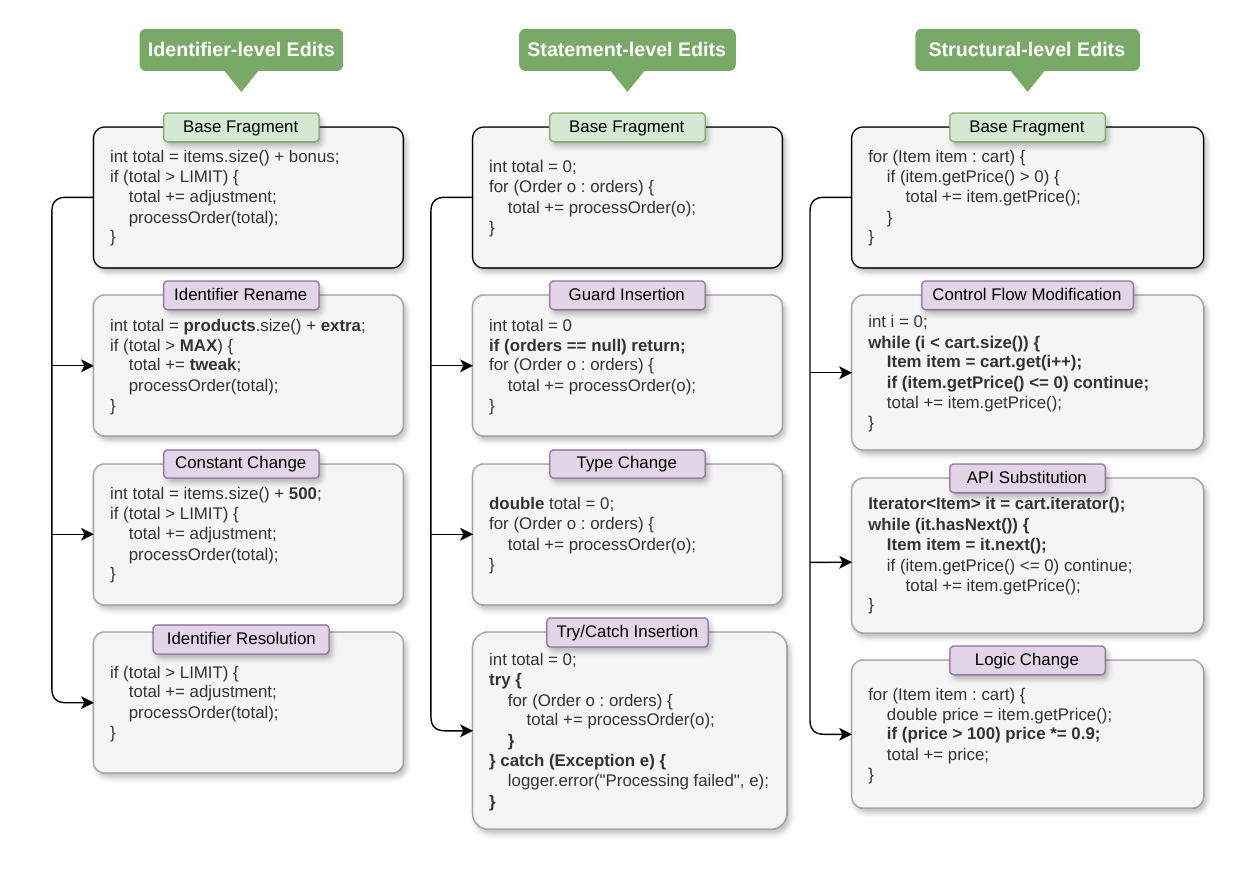}
  \caption{Examples of applying adaptations from the taxonomy in Table~\ref{tab:operators} to a copy-pasted code fragment.}
  \label{fig:taxonomy}
\end{figure*}

Table~\ref{tab:operators} lists the planned mutation operators and Figure~\ref{fig:taxonomy} illustrates each with a concrete example; operators can be applied individually or in combination, with combinations being the primary mechanism for varying complexity. The nine operators span a difficulty gradient: identifier-level edits are expected to be the easiest, structural and semantic edits the hardest, with statement-level edits in between. We will avoid splitting operators that the empirical literature treats as a coherent category (e.g., logic customization is kept as a single catch-all, mirroring its treatment in prior taxonomies \cite{zhang2019analyzing, zhang2026adapteval}).

\begin{table}[t]
  \caption{Planned Adaptation Taxonomy}
  \label{tab:operators}
  \centering
  \small
  \begin{tabularx}{\columnwidth}{l X}
    \toprule
    \textbf{Operator} & \textbf{Description} \\
    \midrule
    \multicolumn{2}{l}{\textit{Identifier-level edits}} \\
    \addlinespace
    Variable rename       & Rename a local identifier to create a conflict or unresolved reference. \\
    \addlinespace
    Constant update       & Change a behaviourally significant literal value. \\
    \addlinespace
    Identifier resolution & Remove a declaration or receiver, leaving an unresolved identifier. \\
    \midrule
    \multicolumn{2}{l}{\textit{Statement-level edits}} \\
    \addlinespace
    Guard insertion       & Add or remove a defensive check. \\
    \addlinespace
    Type change           & Change a local variable type to a non-equivalent alternative. \\
    \addlinespace
    Try/catch edit        & Add or remove an internal \texttt{try/catch} block. \\
    \midrule
    \multicolumn{2}{l}{\textit{Structural and semantic edits}} \\
    \addlinespace
    Control flow modification     & Replace a loop or conditional with a different variant. \\
    \addlinespace
    API substitution      & Replace a library call with a non-equivalent alternative. \\
    \addlinespace
    Logic customization   & Modify an operator, argument, or conditional predicate. \\
    \bottomrule
  \end{tabularx}
\end{table}

\subsection{Mutation Injection Steps}
\label{sec:pipeline}

For each seed fragment F we will generate a mutated version F' as follows:
\begin{enumerate}
  \item Apply one or more mutation operators in reverse to produce F', a fragment that simulates a copied snippet requiring adaptation.
  \item Verify syntactic validity by parsing the output with Spoon; any mutation that produces a parse error is discarded.
  \item Verify that the mutated project still compiles using the Java compiler API (\texttt{javac}); any mutation that produces a compilation error is discarded.
  \item Run the repository test suite against F' via \texttt{mvn test}; any mutation whose output still passes all tests is discarded as trivially reversible, since it provides no measurable adaptation signal.
  \item Record the applied mutations as a JSON metadata file alongside each task, capturing the operator type, the affected AST node identifier, and the original token(s). This record is used during analysis to attribute model successes and failures to specific change types.
\end{enumerate}

We will implement the mutation operators as rewriting passes using Spoon \cite{pawlak2016spoon}, a Java source-to-source transformation framework. Spoon's sniper mode preserves the original formatting of unchanged regions and produces minimal-diff outputs, which is important for re-insertion and for avoiding stylistic artefacts that could bias LLM outputs. Spoon's AST is also used for read-only analysis tasks such as fragment extraction and coverage mapping.

\subsection{Adaptation Task Construction}

Each adaptation task will be constructed as a prompt containing the mutated fragment F' paired with one of three levels of surrounding context (see Figure~\ref{fig:prompt} for the prompt template):

\begin{itemize}
  \item \textbf{C1 -- No context:} the fragment F' is presented alone, with no surrounding code.
  \item \textbf{C2 -- Method context:} the enclosing method body, from its signature (including modifiers, return type, name, and parameter list) to its closing brace.
  \item \textbf{C3 -- File context:} the full source file in which F originally appears, with F' substituted at its original location.
\end{itemize}

Each seed fragment is also assigned a complexity level based on the number of mutations applied. We define four levels: L1 applies a single mutation; L2 applies two mutations from distinct operator families; L3 applies three mutations from at least two distinct families; L4 applies four or more mutations, subject to fragment size and the requirement that the mutated fragment still compiles and fails at least one test. Mutation combinations within each level are sampled to balance operator families where feasible. Tasks for which the framework cannot produce a valid F' at a given complexity level are excluded and reported.

\subsection{Prompt Template}

\begin{figure}[t]
\begin{tcolorbox}[
  title={Prompt Template},
  fonttitle=\small\bfseries,
  colback=gray!5,
  colframe=black!60,
  boxrule=0.4pt,
  left=4pt, right=4pt, top=4pt, bottom=4pt
]
\begin{lstlisting}[style=promptstyle, breakindent=0pt, breakautoindent=false]
<context>
{surrounding context for level C}
</context>

<snippet>
{mutated fragment F'}
</snippet>

Adapt the snippet to fit the surrounding context.
Return only the adapted code, with no explanation.
\end{lstlisting}
\end{tcolorbox}
\caption{Prompt template used across all tasks and context levels. The \texttt{<context>} block is populated according to the context level (C1--C3). No step-by-step instructions or edit lists are provided.}
\label{fig:prompt}
\end{figure}

We will use a temperature of 0.2, which allows limited variance across samples while keeping outputs near each model's high-confidence region; lower temperatures produced near-identical samples in preliminary trials, and higher temperatures increased malformed-output rates. For pass@5 we will draw five independent completions per (model, task, context) condition using the same prompt and decoding settings; a task is counted as pass@5 if any of the five outputs compiles and passes all relevant tests. The five independent completions also serve as a direct measure of output stability under LLM non-determinism~\cite{ouyang2025empirical,kabir2024zs4c}; we will report the variance in pass rates across the five completions per condition as an additional signal. The maximum output length will be set to accommodate the largest fragments in the dataset. We will also record token usage and inference cost per model run to support future cost-benefit analyses.

\section{Research Questions}
\subsection{RQ1: Adaptation Type Difficulty}

\textbf{Which categories of adaptation are hardest for LLMs to perform without explicit instructions?}

Prior work has reported substantial differences in difficulty across adaptation categories even when explicit edit instructions are provided, suggesting that type-level difficulty is a real and measurable phenomenon. We extend this inquiry to the instruction-free setting, where the change types are controlled via mutation injection. Some changes are local, such as renaming variables or updating constants; others are more complex, such as substituting APIs, adding error handling, altering control flow, or modifying conditional logic.

To answer RQ1, we will apply each operator individually to seed fragments (complexity level L1) and evaluate LLM performance separately for each operator category. For each category we will report pass@1 and adaptation-level correctness with 95\% bootstrapped confidence intervals. This analysis will reveal systematic strengths and weaknesses across adaptation types and will directly inform which change types require dedicated model improvements.

\subsection{RQ2: Adaptation Complexity}

\textbf{How does LLM performance change as the number and combination of injected adaptations increases? Does combining individually easy changes produce emergent difficulty beyond what their individual hardness scores would predict?}

Real adaptations rarely involve a single change; a copied fragment may require several coordinated edits at once. No existing benchmark systematically varies the number and combination of simultaneous adaptation changes in a controlled setting. We will address this by varying complexity across levels L1--L4 as defined in Section~\ref{sec:methodology} (Adaptation Task Construction). The L1 single-operator results from RQ1 serve as the baseline for interpreting RQ2: by comparing multi-operator performance at L2--L4 against the individual L1 scores for the same operators, we can isolate emergent difficulty that arises specifically from combining changes rather than from any single operator alone. To avoid confounding the analysis, complexity is capped at L4 (four or more mutations), beyond which the number of simultaneous changes would make it difficult to attribute difficulty to specific operator combinations.

To answer RQ2, we will compare pass@1 and adaptation-level correctness across complexity levels, holding model and context level constant. We will examine whether performance degrades linearly with the number of mutations or whether certain operator combinations create emergent difficulty beyond what their individual difficulty scores would predict.

\subsection{RQ3: Context Granularity}

\textbf{How much surrounding context does an LLM need to perform instruction-free adaptation? Does the optimal level of context depend on the type of adaptation needed?}

Prior work on repository-level code tasks has shown that broader context can improve performance \cite{liu2023repobench, ding2023crosscodeeval}, but it remains open whether context sensitivity varies by adaptation type. This distinction matters for code-reuse tools that must decide how much context to surface.

To answer RQ3, we will evaluate LLM performance across context levels C1, C2, and C3 for each adaptation type and complexity setting. We will model the interaction between context level and adaptation type and report whether certain change types (for example, API substitution versus variable rename) benefit more from broader context than others.

\section{Execution Plan}
\label{sec:execution}
\subsection{Pilot Study}

Before full execution we will run a pilot on a small subset of repositories (3--5 projects, approximately 100 seed fragments) to validate the mutation-injection framework. The pilot must achieve: (i) at least 70\% compile success after mutation, since a lower rate would indicate a systematic problem with operator implementations rather than a dataset property~\cite{wang2025llmmutation}; (ii) at least 60\% of mutated fragments detected as failing by the existing test suite, confirming that mutations produce semantically meaningful changes detectable by the existing tests~\cite{jia2011analysis,wang2025llmmutation}; and (iii) at least 50 valid tasks per operator across the pilot repositories, providing a sufficient minimum for pilot-level feasibility assessment~\cite{efron1993bootstrap,arcuri2011practical}; the exact number of tasks available at higher complexity levels (L2--L4) will depend on how many seed fragments pass through the pipeline at each level and will be reported in the final paper. These thresholds are engineering feasibility checks rather than scientific success criteria; they confirm that the framework is functional before full-scale execution begins. The pilot will additionally include at least 20 L2 combination tasks per operator pair to verify that multi-operator mutations meet the same compile success (70\%) and test-failure (60\%) thresholds as individual operators. If any threshold is not met we will revise the relevant operator implementations or adjust fragment-selection thresholds before proceeding to full execution. Prompt templates and evaluation scripts are finalized during the pilot and locked before full execution begins. The core initial taxonomy, complexity levels, context levels, evaluation metrics, and statistical analysis plan are fixed prior to the pilot and will not be modified; the taxonomy may however be extended via the open-coding phase, with any additions reported in the final paper.

\subsection{Models}

We will evaluate a representative mix of three categories of large language models:
\begin{itemize}
  \item a strong general-purpose model (e.g., GPT-4o~\cite{openai2024gpt4o} or Claude Sonnet 4.6~\cite{anthropic2024claude}), selected as a representative of the current state of the art in general-purpose code understanding;
  \item a state-of-the-art open-source code model (e.g., Qwen3-Coder~\cite{qwen2025qwen3coder}), included to provide a comparison point for openly available code-specialized models;
  \item a reasoning-focused model (e.g., DeepSeek-R1~\cite{deepseek2025deepseekr1}), included to assess whether explicit chain-of-thought reasoning provides an advantage on instruction-free adaptation tasks.
\end{itemize}

Exact version strings will be pinned at inference time and reported in the final paper. All models will be evaluated using the prompt template described in the Mutation Injection Framework (Section~\ref{sec:framework}).

\subsection{Evaluation Metrics}

\textbf{Functional correctness (pass@k).} For each generated adaptation we will re-insert the LLM output into the original location in the codebase and run the repository's existing test suite. A test is considered relevant if it was passing on the unmutated project and covers at least one statement in the seed fragment per JaCoCo. Prior work shows that flaky tests pose a concrete reliability risk in mutation-based evaluation~\cite{shi2019mitigating,luo2014empirical,lam2019idflakies}; we therefore will run each test suite three times on the unmutated project and exclude any test that does not produce consistent results across all runs. The number of runs may be adjusted during the pilot phase depending on the specific repositories. We will report pass@1 and pass@5 following standard practice in code generation benchmarks.

\textbf{Adaptation-level correctness.} For each mutation-injection case we will perform a fine-grained, mutation-by-mutation check. Operator attribution relies on the JSON metadata file recorded at injection time (Section~\ref{sec:pipeline}, step 5), which identifies the operator type and the specific AST node affected; the same node is inspected in the LLM output to determine whether the change was correctly reversed. We will compare the LLM output against the known set of injected mutations and count how many were correctly reversed. When the automatic check is inconclusive---specifically, when the output passes all tests but does not match the target AST at the operator-relevant node---we classify the case as a functional pass with an unresolved adaptation path. A random sample of such cases per operator across models, sized to achieve a 95\% confidence level with a 5\% margin of error, will be reviewed by two independent annotators who judge whether the output constitutes a valid alternative adaptation; inter-rater agreement will be reported using Cohen's kappa.

\subsection{Statistical Analysis}

We will analyze results at three levels corresponding to our research questions. For RQ1 and RQ2, we will compute per-category and per-complexity-level success rates with 95\% confidence intervals obtained via bootstrapping. For RQ3, and for any analysis involving interactions between factors or repeated measures across repositories and fragments, we will use mixed-effects logistic regression with model, adaptation type, context level, and complexity as fixed effects and repository and seed fragment as random effects. This accounts for the binary nature of the pass/fail outcome and the repeated-measures structure of the data. Post-hoc contrasts derived from the fitted model will be used for pairwise comparisons between models and between context levels where needed. All results will be reported with effect sizes and exact p-values. To account for LLM non-determinism~\cite{ouyang2025empirical,kabir2024zs4c}, all pass@1 estimates are derived from the first of the five independent completions drawn per condition; pass@5 uses all five. Variance in pass rates across completions will be reported per model and per operator category.

\section{Threats to Validity}
\label{sec:threats}
\textbf{Synthetic mutations versus real adaptations.} Our test cases rely on programmatically injected mutations. While these mutations are grounded in established taxonomies of how developers adapt copied code \cite{roy2009mutation, zhang2026adapteval, zhang2019analyzing}, they may not perfectly reflect every pattern in natural developer adaptations. We will compare the distribution of injected mutations against real-world distributions reported in prior empirical studies \cite{zhang2026adapteval, zhang2024howdevadapt} and report any categories our operator set does not yet cover. Concretely, we will record the frequency of each injected operator type across our dataset and compare this distribution against the adaptation type frequencies reported in \cite{zhang2019analyzing} and \cite{zhang2024howdevadapt}, which provide detailed empirical counts of real developer adaptations in Java. Any operator types that are over- or under-represented relative to real distributions will be reported explicitly as a limitation of the benchmark.

\textbf{Test coverage limits.} Functional correctness depends on the quality of the repository test suites. We will only use fragments exercised by at least one test case and will report branch and line coverage metrics for all selected seed fragments.

\textbf{Language scope.} The study focuses on Java. Findings may not fully generalize to dynamically typed languages such as Python or JavaScript, where the compiler does not provide an intermediate correctness signal and idioms such as checked exceptions do not exist~\cite{khan2021empirical}. Importantly, the framework itself is language-agnostic in design; Java is the instantiation chosen for this study, and the main language-specific dependency is Spoon, which can be replaced by an equivalent AST framework for other languages. We plan future extensions to additional languages.

\textbf{Build-system heterogeneity.} We restrict the initial dataset to Maven projects, providing a uniform build framework; this restriction may bias the sample toward certain project styles and will be discussed in the final paper. Extending to other build systems is a natural next step requiring no changes to the core framework.

\section{Conclusion}
This registered report proposes a systematic study of instruction-free code fragment adaptation by LLMs. Existing benchmarks do not simultaneously offer instruction-free tasks, fragment-level granularity, controlled change types, and scalability; our mutation-injection framework is designed to provide all four. We will construct thousands of test cases with known ground truth and evaluate LLM performance across adaptation types, complexity levels, and context granularities, yielding insights to guide future code-reuse tools and IDE plugins. The full dataset, mutation framework, and evaluation scripts will be publicly available to support replication and extension.

\bibliographystyle{IEEEtran}
\bibliography{refs}

@inproceedings{treude2017understanding,
  author    = {C. Treude and M. P. Robillard},
  title     = {Understanding {Stack Overflow} Code Fragments},
  year      = {2017},
  booktitle = {2017 IEEE International Conference on Software Maintenance and Evolution (ICSME)},
  pages     = {509--513}
}

@inproceedings{liu2023adaptivepaste,
  author    = {X. Liu and J. Jang and N. Sundaresan and M. Allamanis and A. Svyatkovskiy},
  title     = {{AdaptivePaste}: Intelligent Copy-Paste in {IDE}},
  year      = {2023},
  booktitle = {Proceedings of the 31st ACM Joint European Software Engineering Conference and Symposium on the Foundations of Software Engineering (ESEC/FSE)},
  pages     = {1844--1854}
}

@inproceedings{zhang2026adapteval,
  author    = {T. Zhang and X. Mao and S. Wang and Y. Zhao and Y. Lu and J. Zhang and Z. Zhang and K. Yang and Y. Yu},
  title     = {{AdaptEval}: A Benchmark for Evaluating Large Language Models on Code Snippet Adaptation},
  year      = {2025},
  booktitle = {2025 40th IEEE/ACM International Conference on Automated Software Engineering (ASE)},
  pages     = {1195--1207},
  doi       = {10.1109/ASE63991.2025.00103}
}

@inproceedings{zhang2025instruct,
  author    = {T. Zhang and Y. Yu and X. Mao and S. Wang and K. Yang and Y. Lu and Z. Zhang and Y. Zhao},
  title     = {Instruct or Interact? Exploring and Eliciting {LLMs}' Capability in Code Snippet Adaptation Through Prompt Engineering},
  year      = {2025},
  booktitle = {Proceedings of the IEEE/ACM 47th International Conference on Software Engineering (ICSE)},
  pages     = {566--577}
}

@inproceedings{wang2025wirl,
  author    = {T. Wang and Y. Jiang and C. Dong and Y. Zhang and H. Liu},
  title     = {Wired for Reuse: Automating Context-Aware Code Adaptation in {IDE}s via {LLM}-Based Agent},
  year      = {2025},
  booktitle = {2025 40th IEEE/ACM International Conference on Automated Software Engineering (ASE)},
  pages     = {178--190},
  doi       = {10.1109/ASE63991.2025.00023}
}

@inproceedings{zhang2019analyzing,
  author    = {T. Zhang and D. Yang and C. V. Lopes and M. Kim},
  title     = {Analyzing and Supporting Adaptation of Online Code Examples},
  year      = {2019},
  booktitle = {Proceedings of the 41st International Conference on Software Engineering (ICSE)},
  pages     = {316--327}
}

@article{zhang2024howdevadapt,
  author  = {T. Zhang and Y. Lu and Y. Yu and X. Mao and Y. Zhang and Y. Zhao},
  title   = {How Do Developers Adapt Code Snippets to Their Contexts? An Empirical Study of Context-Based Code Snippet Adaptations},
  year    = {2024},
  journal = {IEEE Transactions on Software Engineering},
  volume  = {50},
  number  = {11},
  pages   = {2712--2731}
}

@inproceedings{terragni2021apization,
  author    = {V. Terragni and P. Salza},
  title     = {{APIzation}: Generating Reusable {APIs} from {StackOverflow} Code Snippets},
  year      = {2021},
  booktitle = {2021 36th IEEE/ACM International Conference on Automated Software Engineering (ASE)},
  pages     = {542--554}
}

@inproceedings{cottrell2008jigsaw,
  author    = {R. Cottrell and R. J. Walker and J. Denzinger},
  title     = {Jigsaw: A Tool for the Small-Scale Reuse of Source Code},
  year      = {2008},
  booktitle = {Companion of the 13th International Conference on Software Engineering (ICSE Companion)},
  pages     = {933--934}
}

@inproceedings{kim2004ethnographic,
  author    = {M. Kim and L. Bergman and T. Lau and D. Notkin},
  title     = {An Ethnographic Study of Copy and Paste Programming Practices in {OOPL}},
  year      = {2004},
  booktitle = {Proceedings of the 2004 International Symposium on Empirical Software Engineering (ISESE)},
  pages     = {83--92}
}

@inproceedings{roy2009mutation,
  author    = {C. K. Roy and J. R. Cordy},
  title     = {A Mutation/Injection-Based Automatic Framework for Evaluating Code Clone Detection Tools},
  year      = {2009},
  booktitle = {2009 International Conference on Software Testing, Verification, and Validation Workshops},
  pages     = {157--166}
}

@article{bellon2007comparison,
  author  = {S. Bellon and R. Koschke and G. Antoniol and J. Krinke and E. Merlo},
  title   = {Comparison and Evaluation of Clone Detection Tools},
  year    = {2007},
  journal = {IEEE Transactions on Software Engineering},
  volume  = {33},
  number  = {9},
  pages   = {577--591}
}

@inproceedings{sajnani2016sourcerercc,
  author    = {H. Sajnani and V. Saini and J. Svajlenko and C. K. Roy and C. V. Lopes},
  title     = {{SourcererCC}: Scaling Code Clone Detection to Big-Code},
  year      = {2016},
  booktitle = {Proceedings of the 38th International Conference on Software Engineering (ICSE)},
  pages     = {1157--1168}
}

@inproceedings{jimenez2024swebench,
  author    = {C. E. Jimenez and J. Yang and A. Wettig and S. Yao and K. Pei and O. Press and K. Narasimhan},
  title     = {{SWE}-bench: Can Language Models Resolve Real-World {GitHub} Issues?},
  year      = {2024},
  booktitle = {The Twelfth International Conference on Learning Representations (ICLR)}
}

@inproceedings{liu2023repobench,
  author    = {T. Liu and C. Xu and J. McAuley},
  title     = {{RepoBench}: Benchmarking Repository-Level Code Auto-Completion Systems},
  year      = {2024},
  booktitle = {The Twelfth International Conference on Learning Representations (ICLR)}
}

@inproceedings{ding2023crosscodeeval,
  author    = {Y. Ding and Z. Wang and W. U. Ahmad and H. Ding and M. Tan and N. Jain and M. K. Ramanathan and R. Nallapati and P. Bhatia and D. Roth and B. Xiang},
  title     = {{CrossCodeEval}: A Diverse and Multilingual Benchmark for Cross-File Code Completion},
  year      = {2023},
  booktitle = {Advances in Neural Information Processing Systems 36 (NeurIPS)}
}

@inproceedings{nguyen2024repoexec,
  author    = {N. L. Hai and D. M. Nguyen and N. D. Q. Bui},
  title     = {On the Impacts of Contexts on Repository-Level Code Generation},
  year      = {2025},
  booktitle = {Findings of the Association for Computational Linguistics: NAACL 2025},
  pages     = {1496--1524},
  doi       = {10.18653/v1/2025.findings-naacl.82}
}

@inproceedings{chi2025editbench,
  author    = {W. Chi and V. Chen and R. Shar and A. Mittal and J. Liang and W. Chiang and A. N. Angelopoulos and I. Stoica and G. Neubig and A. Talwalkar and C. Donahue},
  title     = {{EDIT}-Bench: Evaluating {LLM} Abilities to Perform Real-World Instructed Code Edits},
  year      = {2026},
  booktitle = {The Fourteenth International Conference on Learning Representations (ICLR)}
}

@inproceedings{guo2024codeeditorbench,
  author    = {J. Guo and Z. Li and X. Liu and K. Ma and T. Zheng and Z. Yu and D. Pan and Y. Li and R. Liu and Y. Wang and S. Guo and X. Qu and X. Yue and G. Zhang and W. Chen and J. Fu},
  title     = {{CodeEditorBench}: Evaluating Code Editing Capability of {LLM}s},
  year      = {2025},
  booktitle = {ICLR 2025 Third Workshop on Deep Learning for Code}
}

@article{pawlak2016spoon,
  author  = {R. Pawlak and M. Monperrus and N. Petitprez and C. Noguera and L. Seinturier},
  title   = {{Spoon}: A Library for Implementing Analyses and Transformations of {Java} Source Code},
  year    = {2016},
  journal = {Software: Practice and Experience},
  volume  = {46},
  number  = {9},
  pages   = {1155--1179}
}

@misc{jacoco,
  author       = {{EclEmma Project}},
  title        = {{JaCoCo}: {Java} Code Coverage Library},
  year         = {2009},
  howpublished = {\url{https://www.jacoco.org}}
}

@article{khan2021empirical,
  author  = {F. Khan and B. Adams and D. Varro and S. McIntosh},
  title   = {An Empirical Study of Type-Related Defects in {Python} Projects},
  year    = {2021},
  journal = {IEEE Transactions on Software Engineering},
  volume  = {48},
  number  = {8},
  pages   = {3145--3158}
}

@inproceedings{may2026freshbrew,
  author    = {V. May and D. Misra and Y. Luo and A. Sridhar and J. Gehring and S. S. Ribeiro Junior},
  title     = {{FreshBrew}: A Benchmark for Evaluating {AI} Agents on {Java} Code Migration},
  year      = {2026},
  booktitle = {Proceedings of the 48th International Conference on Software Engineering (ICSE)},
  publisher = {ACM/IEEE}
}

@inproceedings{shi2019mitigating,
  author    = {A. Shi and J. Bell and D. Marinov},
  title     = {Mitigating the Effects of Flaky Tests on Mutation Testing},
  year      = {2019},
  booktitle = {Proceedings of the 28th ACM SIGSOFT International Symposium on Software Testing and Analysis (ISSTA)},
  pages     = {112--122}
}

@inproceedings{luo2014empirical,
  author    = {Q. Luo and F. Hariri and L. Eloussi and D. Marinov},
  title     = {An Empirical Analysis of Flaky Tests},
  year      = {2014},
  booktitle = {Proceedings of the 22nd ACM SIGSOFT International Symposium on Foundations of Software Engineering (FSE)},
  pages     = {643--653}
}

@inproceedings{lam2019idflakies,
  author    = {W. Lam and R. Oei and A. Shi and D. Marinov and T. Xie},
  title     = {{iDFlakies}: A Framework for Detecting and Partially Classifying Flaky Tests},
  year      = {2019},
  booktitle = {Proceedings of the 12th IEEE International Conference on Software Testing, Verification and Validation (ICST)},
  pages     = {312--322}
}

@article{jia2011analysis,
  author  = {Y. Jia and M. Harman},
  title   = {An Analysis and Survey of the Development of Mutation Testing},
  year    = {2011},
  journal = {IEEE Transactions on Software Engineering},
  volume  = {37},
  number  = {5},
  pages   = {649--678}
}

@article{ouyang2025empirical,
  author  = {S. Ouyang and J. M. Zhang and M. Harman and M. Wang},
  title   = {An Empirical Study of the Non-Determinism of {ChatGPT} in Code Generation},
  year    = {2025},
  journal = {ACM Transactions on Software Engineering and Methodology},
  volume  = {34},
  number  = {2},
  pages   = {1--28}
}

@inproceedings{arcuri2011practical,
  author    = {A. Arcuri and L. C. Briand},
  title     = {A Practical Guide for Using Statistical Tests to Assess Randomized Algorithms in Software Engineering},
  year      = {2011},
  booktitle = {Proceedings of the 33rd International Conference on Software Engineering (ICSE)},
  pages     = {1--10}
}

@book{efron1993bootstrap,
  author    = {B. Efron and R. J. Tibshirani},
  title     = {An Introduction to the Bootstrap},
  year      = {1993},
  publisher = {Chapman \& Hall},
  address   = {New York}
}

@article{kabir2024zs4c,
  author     = {A. Kabir and S. Wang and Y. Tian and T. Chen and M. Asaduzzaman and W. Zhang},
  title      = {{ZS4C}: Zero-Shot Synthesis of Compilable Code for Incomplete Code Snippets Using {LLMs}},
  year       = {2025},
  journal    = {ACM Transactions on Software Engineering and Methodology},
  volume     = {34},
  number     = {4},
  articleno  = {90},
  doi        = {10.1145/3702979}
}

@article{wang2025llmmutation,
  author  = {B. Wang and M. Chen and M. Deng and Y. Lin and M. Harman and M. Papadakis and J. M. Zhang},
  title   = {A Comprehensive Study on Large Language Models for Mutation Testing},
  year    = {2026},
  journal = {ACM Transactions on Software Engineering and Methodology},
  note    = {Just Accepted},
  doi     = {10.1145/3805038}
}

@misc{openai2024gpt4o,
  author       = {{OpenAI}},
  title        = {{GPT-4o}},
  year         = {2024},
  howpublished = {\url{https://platform.openai.com/docs/models/gpt-4o}}
}

@misc{anthropic2024claude,
  author       = {{Anthropic}},
  title        = {{Claude} {API}},
  year         = {2024},
  howpublished = {\url{https://www.anthropic.com/api}}
}

@misc{qwen2025qwen3coder,
  author       = {{Qwen Team}},
  title        = {{Qwen3-Coder-480B-A35B-Instruct}},
  year         = {2025},
  howpublished = {\url{https://huggingface.co/Qwen/Qwen3-Coder-480B-A35B-Instruct}}
}

@misc{deepseek2025deepseekr1,
  author       = {{DeepSeek-AI}},
  title        = {{DeepSeek-R1}},
  year         = {2024},
  howpublished = {\url{https://huggingface.co/deepseek-ai/DeepSeek-R1}}
}

\end{document}